Stephen Serjeant, Veronique Buat, Denis Burgarella, Dave Clements, Gianfranco De Zotti, Tomo Goto, Bunyo Hatsukade, Rosalind Hopwood, Narae Hwang, Hanae Inami, Woong-Seob Jeong, Seong Jin Kim, Mirko Krumpe, Myung Gyoon Lee, Matt Malkan, Hideo Matsuhara, Takamitsu Miyaji, Shinki Oyabu, Chris Pearson, Tsutomu Takeuchi, Mattia Vaccari, Ivan Valtchanov, Paul van der Werf, Takehiko Wada, Glenn White

# A new HST/Herschel deep field at the North Ecliptic Pole: preparing the way for JWST, SPICA and Euclid

**Summary**  We propose a co-ordinated multi-observatory survey at the North Ecliptic Pole. This field is the natural extragalactic deep field location for most space observatories (e.g. containing the deepest Planck, WISE and eROSITA data), is in the continuous viewing zones for e.g. Herschel, HST, JWST, and is a natural high-visibility field for the L2 halo orbit of SPICA with deep and wide-field legacy surveys already planned. The field is also a likely deep survey location for the forthcoming Euclid mission. It is already a multi-wavelength legacy field in its own right (e.g. AKARI, LOFAR, SCUBA-2): the outstanding and unparalleled continuous mid-IR photometric coverage in this field *and nowhere else* enables a wide range of galaxy evolution diagnostics unachievable in any other survey field, by spanning the wavelengths of redshifted PAH and silicate features and the peak energy output of AGN hot dust. We argue from the science needs of Euclid and JWST, and from the comparative multiwavelength depths, that the logical approach is (1) a deep (H-UDF) UV/optical tile in the NEP over ~10 square arcminutes, and (2) an overlapping wide-field UV/optical HST survey tier covering >100 square arcminutes, with co-ordinated submm SPIRE mapping up to or beyond the submm point source confusion limit over a wider area and PACS data over the shallower HST tier.

**The natural deep field for JWST, SPICA and Euclid: a new ultra-deep field in the NEP**

The vicinity of the ecliptic poles will always be the natural high visibility / continuous viewing fields for a large class of space observatory missions, e.g. those with L2 halo orbits (e.g. SPICA, Euclid) or low-Earth polar orbits (e.g. HST, AKARI). For any observatory, there is always an obvious efficiency argument for location deep survey fields in regions of highest visibility. The HST Continuous Viewing Zone (CVZ) is ±24° of the ecliptic poles, including e.g. the GOODS-N field, but that of JWST is much more restrictive at just ±5°, and Euclid's deep field locations have the same ±5° restriction (see below). The SEP itself (±5°) is already strongly disfavoured as an extragalactic survey field, with high cirrus backgrounds from the nearby LMC and little ancillary data (not to be confused with the Herschel / AKARI / Spitzer near-SEP field that lies outside the JWST CVZ), but the NEP already has deep imaging fields with AKARI, LOFAR (plus GMRT, eMERLIN, WSRT), Chandra, GALEX, plus shallow Herschel SPIRE data, and several thousand optical/near-IR spectra to date with FLAMINGOS / HECTOSPEC / FMOS.

The proposed Deep-Wide Survey of JWST[1] covers an anticipated ~100 square arcminutes, with many science goals: How did the Hubble sequence form? How did heavy elements form during galaxy evolution? What physical processes drive galaxy evolution? What role does AGN feedback have in galaxy evolution? Answering these questions involves imaging and spectroscopic studies of morphologies, stellar populations, metallicities, dust contents and star formation rates throughout the Hubble time, from optical to mid-infrared wavelengths. Cosmic variance mitigation argues strongly for spreading the total survey area over up to ~10 locations, with the upper limit determined by detector fields of view. Even without considering the NEP's multiwavelength data (in many cases unique to the NEP), observatory efficiency alone argues strongly for at least one JWST survey field of view in the NEP.

*The question is then whether there is an opportunity for existing facilities to generate unique legacy survey products in the NEP that <u>also</u> act as preparatory surveys for JWST.* JWST will not have capabilities below 0.6 microns, so one logical preparatory niche for HST is ultraviolet/optical imaging. The HST CVZ (including the ecliptic poles) includes periods of higher background due to grazing the bright Earth limb[2]. The original HDF-N used this brighter background time for ultraviolet observations, which are not background limited, but the ACS Hubble UDF did not include U-band data and was therefore driven to avoid the Hubble CVZ.

The JWST galaxy evolution science goals place a strong emphasis on the evolution of morphologies, stellar populations and star formation rates *throughout* the Hubble time (*not* just at the highest accessible redshifts), so a natural and obvious objective for HST is to obtain UV morphologies for future JWST-detected galaxies in the JWST CVZ.  The goal of JWST deep-wide survey is to detect SMC-

Stephen Serjeant, Veronique Buat, Denis Burgarella, Dave Clements, Gianfranco De Zotti, Tomo Goto, Bunyo Hatsukade, Rosalind Hopwood, Narae Hwang, Hanae Inami, Woong-Seob Jeong, Seong Jin Kim, Mirko Krumpe, Myung Gyoon Lee, Matt Malkan, Hideo Matsuhara, Takamitsu Miyaji, Shinki Oyabu, Chris Pearson, Tsutomu Takeuchi, Mattia Vaccari, Ivan Valtchanov, Paul van der Werf, Takehiko Wada, Glenn White like galaxies out to z<5, i.e. to AB depths of 30.3 at 0.6 microns (rest-frame ~0.1 microns at z~5). Extending the same ~0.1 micron rest frame continuum sensitivity to lower redshift, we would require AB depths of ~29 at 0.4 microns and ~28 at 0.3 microns, a depth roughly equivalent to the HST UDF at 0.4 microns, and one magnitude deeper than HDF-N at 0.3 microns.

In the shorter term, such a deep optical/UV HST survey would also provide important legacy science that acts as precursors to JWST. The NEP field has the most comprehensive mid-infrared filter coverage of any survey on the sky, covering redshifted PAH and silicate features of starbursts as well as the peak bolometric outputs of AGN-heated dust. The unparalleled 9-filter mid-IR photometry from AKARI gives essentially low-resolution mid-IR spectra of many thousands of galaxies, available nowhere else on the sky, and an ideal precursor data set for both JWST and SPICA. Optical/UV HST data to these depths will test e.g. whether a 'drizzle' of infalling magellanic clouds is responsible for driving ongoing star formation during (and since) the peak of the cosmic star formation history, and will determine how the environment of star-forming galaxies affects the dust composition of their ISM. These cannot be achieved in existing HST deep survey fields.

## Maximising the CVZ targets for JWST: extending GOODS/CANDELS

JWST's targets will not be restricted to sources in its deep-wide 100 square arcminute survey. A shallower tier would provide rarer object populations for targeted follow-ups. The CANDELS survey[3,11] covers 800 square arcminutes to a depth of $H_{AB}<27$, and this wide tier is expected to provide the largest number of z>7 and z>8 galaxies (see figure 1). However, the JWST CVZ is a far more restricted area than that of HST and it is strongly to JWST's advantage to have a large source of targets in its CVZ. A brighter tier at around the CANDELS depths (e.g. $J_{AB}<27$, $H_{AB}<27$) covering at least 100 square arcminutes will be ideal for supplying the largest possible number of targets for JWST spectroscopy in the JWST CVZ. Such a depth is matched to L*; brighter, wider-field survey above L* would reduce in fewer JWST targets for a fixed HST time investment.

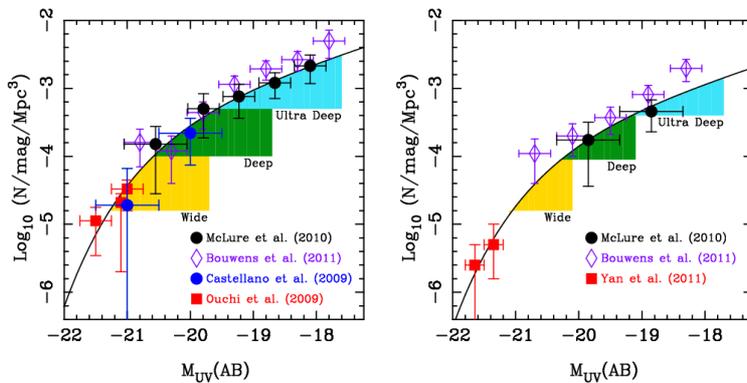

**Figure 1** Galaxy luminosity functions at z>7 (left) and z>8 (right), from [3]. Note the shallow faint-end slope that favours larger-area surveys up to L* for maximising the number of targets.

The natural HST optical/UV complement to such a wide near-infrared survey would be at the GOODS depth, e.g. $i_{AB}\sim27$ in bands BViz, which we argue should be supplemented with U from WFC3, making good use of the HST CVZ bright time (see above) and ensuring at all redshifts that rest-frame ultraviolet measures of the star-forming galaxy morphologies are available for comparison with that of the stellar masses. Prior to JWST and SPICA, the novel advantage this HST survey would have over the existing GOODS/CANDELS-Wide data is the existence of excellent mid-infrared photometric coverage, essentially providing highly-diagnostic low-resolution spectra of redshifted PAH, silicates and AGN. No other field has this advantage of infrared photometric coverage.

Such an HST legacy survey would be very well matched to existing multi-wavelength NEP legacy survey data for intermediate-redshift galaxy evolution studies. Each 100 square arcminutes is expected to contain ~40 massive galaxies with >$10^{11} M_{sun}$ at z~1.5-2.5. This is well matched to the AKARI >3 micron depths: thanks to the negative K-correction on the Rayleigh-Jeans tail of stellar photospheres, the existing multi-wavelength infrared coverage can detect >$10^{11} M_{sun}$ galaxies out to z=4. This HST depth is therefore very well suited to the existing deep ~0.5 deg² AKARI coverage of the NEP. It will be possible to determine the morphological evolution of Milky Way progenitor populations during their main periods of stellar mass assembly. There are already indications that submm-selected

Stephen Serjeant, Veronique Buat, Denis Burgarella, Dave Clements, Gianfranco De Zotti, Tomo Goto, Bunyo Hatsukade, Rosalind Hopwood, Narae Hwang, Hanae Inami, Woong-Seob Jeong, Seong Jin Kim, Mirko Krumpe, Myung Gyoon Lee, Matt Malkan, Hideo Matsuhara, Takamitsu Miyaji, Shinki Oyabu, Chris Pearson, Tsutomu Takeuchi, Mattia Vaccari, Ivan Valtchanov, Paul van der Werf, Takehiko Wada, Glenn White star-forming galaxies in CANDELS have surprisingly quiescent disk morphologies[4], suggesting continuous gas replenishment from minor mergers or cool flows dominates the stellar mass assembly, rather than major mergers. Do we see different star formation modes depending on specific star formation rates, on the relative fractions of obscured and unobscured star formation, on the presence or absence of AGN and/or on the dust composition of the ISM? The comparison of the HST survey data with the unparalleled mid-infrared photometric coverage from AKARI legacy survey data and Herschel (see below) will answer these fundamental questions. Again, these cannot be achieved in existing HST deep survey fields.

**Future Euclid deep field observations: a probable third tier to the CVZ wedding cake**

The ESA Euclid 1.2m space telescope, launching in 2019, will have 40 deg$^2$ of deep fields to visible AB depths of 27.5 and near-IR AB depths of 26 (e.g. J=25.5, Vega). It will also conduct simultaneous near-IR spectroscopy ($\lambda/\Delta\lambda$=250), measuring redshifts for e.g. H$\alpha$ emission-line galaxies. These deep observations are part of the calibration strategy for its 20,000 deg$^2$ dark energy survey, but are in themselves exquisite data sets for galaxy and quasar evolution to z=8 and beyond, and e.g. type Ia supernovae to z~1.5. A major part of the non-dark-energy legacy science of Euclid hinges on these deep fields[5]. The depth and angular resolution of the Euclid data will be ideal for (e.g.) SEDs and high-resolution morphological studies of submm-selected galaxies and their stellar populations (including stellar ages and masses and dust extinction via panchromatic SED fits), and the Euclid deep survey data will be a legacy goldmine for which many science areas will benefit from the far-IR perspective. However, orbital constraints strongly restrict the field choices to within 5 degrees of the ecliptic poles.

Although the division of these 40 deg$^2$ between the two poles is still to be finalised, the available or planned multiwavelength data is already weighted very strongly in favour of the North Ecliptic Pole (see above). In contrast, the potential Euclid coverage of the South Ecliptic Pole would miss *all* the currently available nearby multi-wavelength extragalactic survey data. All other famous extragalactic survey fields are also very difficult to incorporate in Euclid's scanning strategy. The ecliptic poles will *always* be the natural deep field locations for many satellite survey missions, already including the deepest Planck, WISE and eROSITA coverage. The existence in Euclid's main calibration field of wide-field HST survey data, deeper than Euclid, will be useful for the Euclid performance verification.

**Herschel observations of the NEP**

The NEP currently has a single scan 9 deg$^2$ SPIRE map in the NEP. SPICA will have no submm mapping capability, so as things stand, when the Euclid data comes it is quite plausible that the extragalactic community will regret that at least one Euclid deep field was not better-studied by Herschel while the opportunity existed.

We therefore proposed deeper mapping of the NEP field in the "must-do" observations call earlier this year, to reach the SPIRE confusion limit (3 further repeats of the 9 deg$^2$ SPIRE AOR already conducted in the NEP, at any available orientations). SPIRE is extremely efficient so these observations are relatively cheap, at <20 hours. By reaching the confusion limit we maximise the information available on individual galaxy populations detected by the two HST tiers above.

The Euclid deep survey and SPIRE depths are quite well-matched for submm-*selected* galaxies. In order to detect submm galaxies securely up to z=4 and beyond, depths of around K(Vega)=22 are needed[6]. The J–K(Vega) colours of submm galaxies[7,8] are around 1 to 3, so a survey depth of J~25 (Vega) is needed for secure detection of submm galaxies. The anticipated survey depth of the Euclid deep survey is J~25.5 (Vega, 5$\sigma$ point source), i.e. well-suited to the detection of SPIRE galaxies.

Herschel will also be indispensible for IR-emitting galaxies in general. Euclid's deep imaging and spectroscopic survey will in total detect 10$^5$ galaxies at 1.2<z<2 (the ground-based "redshift desert") with H$\alpha$ and [OIII] detections from Euclid's near-IR spectroscopy, thousands of galaxies at z>6, several tens of z>8 galaxies and quasars, ~3000 type Ia SNe to z~1.2 with near-IR light curves and colours, and much more besides[5]. Integrating to the SPIRE confusion limit maximises the available submm photometric information on each object (or its host galaxy) in all these populations. Euclid's 0.2'' resolution will also detect e.g. strong gravitational lenses, already known to exist in significant numbers in Herschel surveys[9,10]. Combining Euclid's imaging / spectroscopy deep survey with

Stephen Serjeant, Veronique Buat, Denis Burgarella, Dave Clements, Gianfranco De Zotti, Tomo Goto, Bunyo Hatsukade, Rosalind Hopwood, Narae Hwang, Hanae Inami, Woong-Seob Jeong, Seong Jin Kim, Mirko Krumpe, Myung Gyoon Lee, Matt Malkan, Hideo Matsuhara, Takamitsu Miyaji, Shinki Oyabu, Chris Pearson, Tsutomu Takeuchi, Mattia Vaccari, Ivan Valtchanov, Paul van der Werf, Takehiko Wada, Glenn White

Herschel will determine the evolving obscuration to star formation throughout the history of the Universe, connect AGN host galaxy star formation with morphologies (Euclid will resolve morphologies of all AGN hosts), and constrain cosmic dust generation at the highest accessible redshifts.

The combination of comprehensive infrared data, SPIRE submm data and the HST wider tier data, together creates a strong near-term driver for PACS 100-160 micron data in the deeper central field covered by the HST CANDELS-depth survey discussed above, in order to cover the wavelengths of the peak bolometric outputs of star-forming galaxies prior to SPICA. This enables robust estimates of masses of cold dust phases in the ISM and of total bolometric outputs of GMCs, as well as characterising the SED evolution as a function of galaxy morphology (from HST) and of PAH dust composition in the ISM (from AKARI). A 500 micron-detected galaxy at z=5 at the submm confusion limit would have 100 (160) micron fluxes of approximately 5 (10) mJy.

**Other international legacy considerations**

The L2 halo orbit planned for SPICA makes it very likely that the ecliptic poles will have excellent visibility for SPICA. The Japanese community already heavily invested in the AKARI NEP field, and have already determined that the NEP will also be the target of deep and wide-field SPICA legacy surveys (Wada, priv. comm.). SPICA will be equipped with slit-less multi-object spectroscopy; multi-roll angle observations, essential to solve the degeneracy of spectrum overlap, can be realized only in the NEP regions. While there is value in placing new deep fields in areas accessible to ALMA, the NEP is ideally placed for NOEMA, which will have sensitivities within a factor of two or three of ALMA. JWST and Euclid will eventually provide deep imaging in the 0.6-1 micron range, but in the meantime our recommendation is that our two suggested HST NEP survey tiers also provide photometric coverage in this wavelength range. The NEP field has a slightly higher density of foreground stars than other HST fields, but this brings benefits for AO: foreground sources are necessary for AO observations, even in case of the artificial laser guide stars, because a natural guide star is needed for a tip-tilt system.

**Conclusions**

The NEP CVZ is the natural location for JWST and SPICA deep fields, and will always be the only natural deep field location for a wide class of space observatories. It is also already a legacy survey field in its own right, with a unique and comprehensive set of mid-infrared data characterising the PAH and silicates in the galaxies' ISM and their dust-shrouded AGN. We propose two tiers of HST survey and a further coordinated Herschel survey, both in support of the unique existing legacy survey data with science goals that cannot be achieved in any other field, and in preparation for JWST and SPICA:

(a) an ultra-deep H-UDF-like survey, extended to U-band, in the expectation of locating ~10% of the proposed deep-wide 100 square arcminute JWST survey in the NEP;
(b) a shallow GOODS/CANDELS depth survey covering >100 square arcminutes, to maximise the follow-up targets in the JWST CVZ;
(c) Herschel PACS coverage of tier (b)
(d) Herschel SPIRE coverage of a wider 9 deg$^2$ field to the point source confusion limit (also proposed through the "must-do" Director's time opportunity), in support of forthcoming Euclid data as well as of existing multi-wavelength survey data.